\documentclass[
12pt,
amsmath,
superscriptaddress,
floatfix,
noshowkeyws,
notitlepage,
noshowpacs,
]{revtex4-1}
\usepackage{graphicx}
\usepackage{amsmath}
\usepackage{comment}
\usepackage{amssymb}

\usepackage{subfigure}

\begin{document}

\title{Casimir force in dense confined electrolytes}

\author{Alpha A. Lee}
\affiliation{Cavendish Laboratory, University of Cambridge, Cambridge CB3 0HE, United Kingdom}

\author{Jean-Pierre Hansen}
\affiliation{Department of Chemistry, University of Cambridge, Cambridge, CB2 1EW, United Kingdom}

\author{Olivier Bernard}
\affiliation{Sorbonne Universit\'e, CNRS, Physicochimie des \'electrolytes et nanosyst\`emes interfaciaux, UMR PHENIX, F-75005, Paris, France}

\author{Benjamin Rotenberg}
\email{benjamin.rotenberg@sorbonne-universite.fr}
\affiliation{Sorbonne Universit\'e, CNRS, Physicochimie des \'electrolytes et nanosyst\`emes interfaciaux, UMR PHENIX, F-75005, Paris, France}
\affiliation{R\'eseau sur le Stockage Electrochimique de l'Energie (RS2E), FR CNRS 3459, 80039 Amiens Cedex, France}

\begin{abstract}
Understanding the force between charged surfaces immersed in an electrolyte
solution is a classic problem in soft matter and liquid-state theory. Recent
experiments showed that the force decays exponentially but the characteristic
decay length in a concentrated electrolyte is significantly larger than what
liquid-state theories predict based on analysing correlation functions in the
bulk electrolyte. Inspired by the classical Casimir effect, we consider an
alternative mechanism for force generation, namely the confinement of density
fluctuations in the electrolyte by the walls. We show analytically within the
random phase approximation,
which assumes the ions to be point charges, 
that this fluctuation-induced force is attractive and also decays
exponentially, albeit with a decay length that is half of the bulk correlation
length. 
These predictions change dramatically when excluded volume effects are accounted
for within the mean spherical approximation. At high ion concentrations the
Casimir force is found to be exponentially damped oscillatory as a function of the
distance between the confining surfaces.
Our analysis does not resolve the riddle of the anomalously long
screening length observed in experiments, but suggests that the
Casimir force due to mode restriction in density fluctuations could be an
hitherto under-appreciated source of surface-surface interaction. 
\end{abstract}

\makeatother
\maketitle


\section{Introduction} 
Understanding the structure, phase behaviour and dynamics of ionic liquids and
concentrated electrolytes, both in the bulk and near interfaces, is a
longstanding challenge. Since the pioneering work of Helmholtz
\cite{helmholtz1853}, Debye and H\"{u}ckel \cite{debye1923}, Onsager
\cite{onsager1934deviations} and many others, much recent progress has been made
using the statistical mechanics tools of the theory of classical liquids
\cite{hansen2013theory}. A large body of ``exact" results and sum-rules was
established \cite{martin1988sum}, while the Ornstein-Zernike (OZ) formalism
\cite{ornstein1914accidental} and classical density functional theory
\cite{percus1962approximation} became the basis of numerous approximate theories
of the structure, including non-linear integral equations for the pair correlation functions \cite{hansen2013theory}; amongst these theories the mean spherical approximation (MSA) plays an important role, since it allows for analytic solutions of simple, semi-realistic models of ionic liquids \cite{waisman1972mean,waisman1972mean1}, which will be used in the present paper. The OZ formalism was also put to good use to examine the asymptotic decay of pair correlation functions and density profiles at interfaces \cite{attard1993asymptotic,leote1994decay,attard1996electrolytes}. Although different analytical or numerical theories predict different dependences of the correlation (or screening) length on ion concentration, the theoretical predictions converge on two qualitative features: (1) the decay of the correlation is exponential; and (2) the longest correlation length in a concentrated electrolyte is of the same order of magnitude as the (mean) ion diameter. 

However, recent experiments suggest an ``underscreening" phenomenon, namely the
existence of an anomalously large decay length which is incongruent with the
above mentioned theoretical predictions
\cite{gebbie2013ionic,gebbie2015long,smith2016electrostatic}. Surface force
balance experiments reveal hat the force acting between negatively charged mica
surfaces immersed in an electrolyte decays exponentially with surface separation $L$, but the decay (or screening) length $\lambda_S$ scales as \cite{lee2017scaling}: 
\begin{equation}
\frac{\lambda_S}{\lambda_D} \sim  
\begin{cases} 
1,  & a/\lambda_D \ll 1 \\ 
(a/\lambda_D)^3, & a/\lambda_D \gg 1,  
\end{cases} 
\label{scaling_exp}
\end{equation} 
with 
\begin{equation}
\lambda_D = \frac{1}{\sqrt{4 \pi l_B (\rho_+z_+^2 + \rho_- z_-^2) }}
\end{equation} 
the Debye length, $\rho_\pm$ ($z_\pm$) the number density (valence) of the
cations/anions, $a$ the ionic radius, while $l_B = e^2/(4 \pi \epsilon
\epsilon_0 k_B T)$ is the Bjerrum length with $\epsilon$ the dielectric constant
of the electrolyte, which depends on ion concentration This scaling relation has
been verified for various electrolyte chemistries, ranging from pure ionic
liquids (e.g. room temperature molten salts) and ionic liquid-organic solvent
mixtures to aqueous alkali halide solutions. A scaling theory has been proposed,
based on identifying solvent molecules as effective charge carriers, with an effective charge determined by thermal fluctuations \cite{lee2017scaling,lee2017underscreening}. More recently, a first-principles analysis based on Landau fluctuation theory and the MSA has been put forward, which confirms that $\lambda_S/\lambda_D$ has a power law dependence on $a/\lambda_D$, albeit with a considerably smaller exponent compared to the experimental findings summarised in Equation (\ref{scaling_exp})  \cite{rotenberg2017underscreening}. 

In this paper, we explore an additional mechanism of force generation in confined systems, namely the classical counterpart of the celebrated quantum Casimir effect of an electromagnetic field fluctuation-induced force acting between the confining surfaces \cite{casimir1948attraction}. The classical Casimir effect is observed in high temperature confined systems, where the thermal fluctuations now play the role of quantum field fluctuations. Restrictions on the possible Fourier components (or modes) of thermal fluctuations imposed by spatial confinement generate the classical Casimir force. Large amplitude critical fluctuations in a fluid close to a thermodynamic critical point strongly enhance the classical Casimir effect, where the universality of critical scaling laws entails a corresponding universality of the Casimir force \cite{fisher1978} (for a recent review of the classical Casimir force, see \cite{gambassi2009casimir}). 

We examine the possibility of an observable Casimir force in confined ionic fluids under conditions inspired by the aforementioned experimental setups \cite{gebbie2013ionic,gebbie2015long,smith2016electrostatic}. No critical fluctuations are involved, but the infinite range of the Coulombic interactions is expected to significantly affect the resulting Casimir force. This question has already been explored in the high temperature limit within Debye-H\"{u}ckel theory of point ions, for a variety of boundary conditions, and using a microscopic description of the confining metallic or dielectric media \cite{jancovici2004screening,jancovici2005casimir,buenzli2005casimir,hoye2009casimir}. 

This paper describes an attempt to go beyond the point ion description by
considering finite size ions to account for excluded volume effects which are
crucial for concentrated electrolytes. In Section~\ref{sec:freeenergy}, we first consider the
point ion limit using a systematic approach inspired by a paper dealing with
Casimir force in confined non-equilibrium systems \cite{brito2007generalized},
while excluded volume effects are included within the MSA in
Section~\ref{sec:msa}. Some
concluding remarks are made in Section~\ref{sec:conclusion}.  

\section{The free energy of fluctuation modes}
\label{sec:freeenergy} 

We begin our analysis by expressing the free energy in terms of fluctuation modes. Let $F = F(\rho_+,\rho_-)$ be the free energy of a bulk electrolyte with cation density $\rho_+$ and anion density $\rho_-$. We expand around the mean density, i.e. $\rho_\alpha = \rho_\alpha^{0} + \delta \rho_\alpha$, and write 
\begin{equation}
F = F(\rho_+^{0},\rho_-^{0}) + \sum_{\alpha=\pm} \delta \rho_\alpha \frac{\partial F}{\partial \rho_\alpha} \Bigg|_{\rho_+^{0},\rho_-^{0}}+ \frac{1}{2} \sum_{\alpha,\beta =\pm }  \delta \rho_\alpha \delta \rho_\beta \frac{\partial^2 F}{\partial \rho_\alpha \partial \rho_\beta} \Bigg|_{\rho_+^{0},\rho_-^{0}}
\end{equation} 
Defining $\Delta F = F -F(\rho_+^{0},\rho_-^{0})$, and noting that $\left<
\delta \rho_\alpha \right> = 0$, where $\left< \cdot\right>$ denotes thermal average, we obtain 
\begin{equation}
\Delta F =   \frac{1}{2} \sum_{\alpha,\beta = \pm}  \left< \delta \rho_\alpha \delta \rho_\beta \right> \frac{\partial^2 F}{\partial \rho_\alpha \partial \rho_\beta}\Bigg|_{c_+^{0},c_-^{0}} = \frac{1}{2} \sum_{\alpha,\beta = \pm}  \left< \delta \rho_\alpha \delta \rho_\beta \right> \chi^{-1}_{\alpha \beta}
\end{equation} 
where we have defined the partial response functions \cite{rotenberg2017underscreening}
\begin{equation}
\frac{\partial^2 F}{\partial \rho_\alpha \partial \rho_\beta} \Bigg|_{\rho_+^{0},\rho_-^{0}} = \chi_{\alpha \beta}^{-1}. 
\label{part_res}
\end{equation} 
We can express the fluctuations in terms of Fourier modes $ \delta \rho_\alpha(
\mathbf{r}) = \frac{1}{V} \sum_\mathbf{k} e^{- i \mathbf{k} \cdot \mathbf{r}}
\delta \rho_{\alpha,\mathbf{k}} $, and the correlations of the fluctuations are
related to the structure factors $S_{\alpha \beta}(\mathbf{k}) = \left< \delta
\rho_{\alpha,\mathbf{k}} \delta \rho_{\beta,\mathbf{-k}} \right>/V$, which are in principle experimentally measurable using techniques such as neutron scattering. 

We now consider an electrolyte solution confined between two infinite charged
walls separated by a distance $L$. For a strongly charged surface, one might
imagine that the concentration fields of the cations and anions are pinned on
the surface, or at the very least the surface anchors the fields and
significantly reduces the magnitude of fluctuations. Assuming that the fields
are pinned at the walls (\emph{i.e.} $\delta \rho_\alpha = 0$ at the walls), the wavenumber of the fluctuation modes normal to the surfaces can only take discrete values $k_n = n \pi/L$. Therefore, the fluctuation energy inside the slit is given by 
\begin{equation}
\Delta F_{\mathrm{in}} =  \int \frac{\mathrm{d}^2 \mathbf{k}}{(2\pi)^2} \left[ \frac{\pi}{L} \sum_{n = 1}^{\infty} \sum_{\alpha,\beta = \pm} \chi^{-1}_{\alpha \beta} S_{\alpha \beta}\left(\sqrt{k^2 + \left(\frac{n \pi}{L}\right)^2 } \right)  - \int_{0}^{\infty} \mathrm{d}p \sum_{\alpha,\beta = \pm} \chi^{-1}_{\alpha \beta} S_{\alpha \beta}\left(\sqrt{k^2 + p^2 } \right)\right] 
\label{fluct_energy}
\end{equation} 
where we have subtracted the energy in the limit when $L \rightarrow \infty$,
and exploited the symmetry of the summand and integrand with respect to negative
$n$ and $p$. We note that the $n=0$ term is irrelevant since it is independent of $L$. The resulting Casimir force is simply the derivative of the fluctuation energy with respect to the surface separation 
\begin{equation} 
f_{\mathrm{Casimir}} = - \frac{\partial \Delta F_{\mathrm{in}} }{\partial L}. 
\label{casimir_force1}
\end{equation}
Note that we have implicitly assumed that charged surfaces do not affect the structure factors $S_{\alpha \beta}(k)$ -- this assumption restricts the validity of our analysis to the far field limit when the walls are far apart. 

Equations (\ref{fluct_energy})-(\ref{casimir_force1}) relate the bulk response
functions and the structure factors to the Casimir force. We next turn to
estimating those quantities for a two-component electrolyte. Following ref.~\cite{rotenberg2017underscreening}, we introduce the wavenumber-dependent partial response functions $\hat{\chi}_{\alpha \beta}$, defined by
\begin{equation}
\hat{\chi}^{-1}_{\alpha \beta}(k) = \frac{\delta_{\alpha \beta}}{\rho_\alpha} - \hat{c}_{\alpha \beta} (k),  
\label{k_susceptibility}
\end{equation} 
where $\hat{c}_{\alpha \beta} (k)$ is the Fourier transform of the
OZ direct correlation function. Using the definition of the structure factor in terms of the total correlation function $\hat{h}_{\alpha \beta}(k)$
\begin{equation} 
S_{\alpha \beta}(k) = \rho_\alpha \delta_{\alpha \beta} + \rho_\alpha \rho_\beta \hat{h}_{\alpha \beta}(k)
\end{equation}
it can be shown \cite{hansen2013theory,rotenberg2017underscreening} that 
\begin{equation} 
\begin{pmatrix}
    S_{++} & S_{+-} \\
   S_{-+} & S_{--} 
  \end{pmatrix} = \frac{1}{\hat{\chi}^{-1}_{++} \hat{\chi}^{-1}_{--} - \hat{\chi}^{-1}_{+-}\hat{\chi}^{-1}_{-+}} 
  \begin{pmatrix}
    \hat{\chi}^{-1}_{++} & -\hat{\chi}^{-1}_{+-} \\
   -\hat{\chi}^{-1}_{-+} & \hat{\chi}^{-1}_{--} 
  \end{pmatrix}. 
\label{struct_fact_matrix}
\end{equation}
To make further progress, we can split the direct correlation functions into the Coulomb
part and the short-range part:
\begin{equation}
\hat{c}_{\alpha \beta}(k) = -\frac{4 \pi z_\alpha z_\beta l_B}{k^2} + \hat{c}^{s}_{\alpha \beta}(k). 
\label{decomp}
\end{equation} 
The Random Phase Approximation (RPA) assumes that $\hat{c}^{s}_{\alpha \beta}(k)=0$, and in this limit Equation (\ref{k_susceptibility}) can be substituted into Equation (\ref{struct_fact_matrix}) to yield analytical expressions for the structure factors. 

The partial response functions, Equation (\ref{part_res}), can be evaluated by noting that the free energy density of an electrolyte in the random phase approximation reads
\begin{equation}
\frac{F}{k_B T V}=   \rho_+ \left[ \log(a^3 \rho_+) -1\right] +  \rho_- \left[ \log(a^3 \rho_-) -1 \right]
- \frac{1}{12 \pi \lambda_D^3} . 
\end{equation}
Taking derivatives with respect to $\rho_+$ and $\rho_-$, we thus arrive at 
\begin{equation}
\chi^{-1}_{\alpha \beta} = V k_B T \left( \frac{\delta_{\alpha \beta}}{\rho_\alpha} -  \pi  \lambda_D l_B^2 z^2_\alpha z^2_\beta \right)
\; . 
\end{equation} 
For a $1:1$ electrolyte, $z_+ =-z_- = 1$, $\rho_+ =\rho_- = \rho/2$, and the sum of structure factors can be written as  
\begin{eqnarray}
\sum_{\alpha,\beta = \pm} \chi^{-1}_{\alpha \beta} S_{\alpha \beta}\left(k \right) &= & \frac{V k_B T}{2} \left[ \left(2 - \frac{l_B}{4 \lambda_D}\right) \frac{2 \lambda_D^2 k^2+1}{\lambda_D^2 k^2+1} - \frac{l_B}{4 \lambda_D} \frac{1}{1 + \lambda_D^2 k^2}   \right] \nonumber \\ 
& = & V k_B T \left( 2- \frac{l_B}{4 \lambda_D} - \frac{1}{1+\lambda_D^2 k^2} \right)
\label{structure_fact}
\end{eqnarray}
we first note that the constant term drops out of the Casimir force as the sum and the integral cancel out, 
\begin{equation}
\sum_{n = 1}^{\infty}   \frac{\pi}{L}  - \int_{0}^{\infty} \mathrm{d}p =   \sum_{n = 1}^{\infty} \left(\frac{\pi}{L}  -  \int_{\frac{(n-1) \pi}{L}}^{\frac{n \pi}{L}} \mathrm{d}p\right) = 0. 
\label{const_term}
\end{equation}
The crucial step of our analysis is to note that 
\begin{equation}
\frac{\pi}{L} \sum_{n = 1}^{\infty} \frac{1}{1+\lambda_D^2 k^2 +\lambda_D^2 \left( \frac{n \pi}{L}\right)^2} = \frac{\pi}{L} \frac{1}{2 (1 + \lambda_D^2 k^2 )} \left[ \frac{L}{\lambda_D} \sqrt{1 + \lambda_D^2 k^2} \coth \left(  \frac{L}{\lambda_D} \sqrt{1 + \lambda_D^2 k^2} \right) - 1 \right] 
\label{sum_discrete}
\end{equation}
and 
\begin{equation}
\int_{0}^{\infty} \mathrm{d}p \frac{1}{1+\lambda_D^2 k^2 + \lambda_D^2 p^2} = \frac{\pi}{2 \lambda_D \sqrt{1 + \lambda_D^2 k^2} },
\label{sum_integral}
\end{equation}
Substituting 
the difference between Equations (\ref{sum_discrete}) and (\ref{sum_integral})
into Equation (\ref{fluct_energy}), 
and multiplying by $L$, we obtain the free energy per unit area (instead of
volume):
\begin{equation}
\frac{\Delta F_{\mathrm{in}}}{A k_B T}  = \frac{1}{4} \left( \int_0^{\infty} \frac{k}{1+\lambda_D^2 k^2} \mathrm{d}k - L \int_0^{\infty}  k \frac{\coth \left( \frac{L}{\lambda_D} \sqrt{1+ k^2 \lambda_D^2 }\right) -1}{\lambda_D \sqrt{1 + \lambda_D^2 k^2} } \mathrm{d}k \right), 
\label{RPA_calculation}
\end{equation}
with $A$ the plate area.
While the first term of Equation (\ref{RPA_calculation}) diverges logarithmically,
it is $L$-independent and therefore does not contribute to the disjoining force. 
The second term can be integrated analytically to give 
\begin{equation}
\frac{\Delta F_{\mathrm{in}}}{A k_B T}  = - \frac{1}{4} \left[
\frac{L}{\lambda_D^3} - \frac{1}{\lambda_D^2} \log\left( 2 \sinh
\frac{L}{\lambda_D} \right)  \right]. 
\end{equation}
Therefore, the Casimir force per unit area is 
\begin{equation} 
\frac{f_{\mathrm{Casimir}}}{A} =  \frac{k_B T}{4\lambda_D^3} \left(1- \coth \frac{L}{\lambda_D} \right). 
\label{casimir_force}
\end{equation}
Perhaps surprisingly, Equation (\ref{casimir_force}) reveals that the Casimir force is \emph{attractive}, and has an asymptotic decay length of of $\lambda_D/2$.

\section{Hard core repulsion and the Mean-Spherical Approximation}
\label{sec:msa}
 
The RPA ignores hard-core interactions and assumes point-like ions. This
approximation is unreasonable in dense ionic systems such as ionic liquids and
concentrated electrolytes. To include hard-core interactions, we consider the
Mean Spherical Approximation (MSA). The MSA direct correlation function for a
two component hard sphere electrolyte with cations and anions having equal
diameters $\sigma$ has been derived in pioneering papers
\cite{wertheim1963exact,thiele_equation_1963,waisman1972mean,waisman1972mean1}, and reads  
\begin{align}
\hat{c}^{s}_{\alpha\beta}(k) &= \frac{4 \pi \sigma^3}{(k\sigma)^6} \left[ 24
d_{\alpha\beta} - 2 b_{\alpha\beta} (k\sigma)^2 + e_{\alpha\beta} (k\sigma)^4  \right. \nonumber \\
& - \left \{ 24 d_{\alpha\beta} - 2(b_{\alpha\beta} +6 d_{\alpha\beta})(k\sigma)^2 
+(a_{\alpha\beta} +b_{\alpha\beta} +d_{\alpha\beta} +e_{\alpha\beta}
)(k\sigma)^4 \right \} \cos (k\sigma) \nonumber \\
 & \left. + \left \{ -24d_{\alpha\beta} (k\sigma) + (a_{\alpha\beta}+2
b_{\alpha\beta}+4 d_{\alpha\beta})(k\sigma)^3 \right \} \sin (k\sigma)\right] 
\label{MSA_dca}
\end{align} 
where
\begin{equation*} 
a_{\alpha\beta} = - \frac{(1 + 2 \eta)^2}{(1-\eta)^4} - 2
B\left(\frac{\sigma}{\lambda_D}\right) \frac{l_B}{\sigma} z_\alpha z_\beta, 
\end{equation*} 
\begin{equation*} 
b_{\alpha\beta} = - \frac{6 \eta (1 + \frac{\eta}{2})^2}{(1-\eta)^4} + \left[
B\left(\frac{\sigma}{\lambda_D}\right) \right]^2 \frac{l_B}{\sigma} z_\alpha z_\beta, 
\end{equation*} 
\begin{equation*} 
d_{\alpha\beta} =  - \frac{ \eta (1 + 2 \eta)^2}{2 (1-\eta)^4}, 
\end{equation*} 
\begin{equation*} 
e_{\alpha\beta} =  \frac{l_B}{\sigma} z_\alpha z_\beta, 
\end{equation*} 
\begin{equation*} 
B(x) =  \frac{x^2 + x - x \sqrt{1+ 2x} }{x^2},
\end{equation*} 
with $\eta = (\pi/6)\sum_\alpha \rho_\alpha \sigma^3$ the total packing
fraction. Substituting Equation (\ref{MSA_dca}) into Equation (\ref{decomp})
yields the full direct correlation function. Unlike the RPA, the hard core
repulsion causes the MSA structure factor to be oscillatory and to decay to zero in the $k \rightarrow \infty$ limit. 

To proceed further, we first evaluate numerically the difference between the sum and the integral 
\begin{equation}
G_{\alpha\beta}(k_{\parallel}, L) = \frac{\pi}{L} \left( \sum_{n=1}^{\infty} S_{\alpha\beta}\left(\sqrt{k_{\parallel}^2 + \frac{n \pi}{L}} \right) - \int_0^{\infty} \; S_{\alpha\beta}\left(\sqrt{k_{\parallel}^2 + \frac{n \pi}{L} } \right) \mathrm{d}n \right). 
\label{sum_difference}
\end{equation}
and note that both the sum and the integral are convergent since the structure
factors decay asymptotically as: 
\begin{equation}
\sigma^3 S_{\alpha\beta}(k) \sim \frac{3 \eta}{\pi} \delta_{\alpha \beta} - \frac{36 \eta^2}{
\pi} (a_{\alpha \beta} + b_{\alpha \beta}+d_{\alpha \beta} + e_{\alpha \beta})
\frac{\cos (k\sigma) }{(k\sigma)^2}, \; \mathrm{when} \; k \rightarrow \infty. 
\end{equation}
and we showed in Equation (\ref{const_term}) that a constant term has no bearing
on the Casimir force and can be ignored. Using the Euler-Maclaurin formula, we
can expand $G_{\alpha\beta}(k_{\parallel}, L)$ asymptotically in $1/L$: 
\begin{equation} 
G_{\alpha\beta}(k_{\parallel}, L) = - \frac{1}{2} \frac{\pi}{L} S_{\alpha\beta}\left(k_{\parallel} \right) + o(L^{-1}). 
\label{EM_expansion}
\end{equation}  
Since the relevant quantity is the Casimir energy per unit area, we need to
multiply Equation (\ref{sum_difference}) by $L$ at the end of the calculation,
such that the first term in (\ref{EM_expansion}) becomes actually a (diverging) constant independent of $L$ (c.f. the first term in Equation (\ref{RPA_calculation})). As such, we must subtract it before numerically integrating over $k_{\parallel}$. All in all, the Casimir energy (per unit volume) reads 
\begin{equation}
E_{\mathrm{Casimir}}(L) = \sum_{\alpha \beta} \chi_{\alpha \beta}^{-1} F_{\alpha \beta}(L)
\label{casimir_energy_msa}
\end{equation}
where 
\begin{equation} 
F_{\alpha \beta}(L) =\frac{1}{(2 \pi)^2} \int_0^{\infty}2 \pi   k_{\parallel}
\left[  G_{\alpha\beta}(k_{\parallel}, L) + \frac{1}{2L}
S_{\alpha\beta}\left(k_{\parallel} \right) \right] \mathrm{d}k_{\parallel} 
\label{decay_integrate}
\end{equation} 
and 
\begin{align}
\chi^{-1}_{\alpha \beta} &= \chi^{-1}_{\alpha \beta, RPA} + \hat{c}^{s}_{\alpha \beta}(0) \nonumber \\ 
& = V k_B T \left[  \frac{\delta_{\alpha \beta}}{\rho_\alpha} -  \pi  \lambda_D
l_B^2 z^2_\alpha z^2_\beta - \frac{\pi}{3} \sigma^3 (4 a_{\alpha \beta} + 3
b_{\alpha \beta} + 2 d_{\alpha \beta} + 6 e_{\alpha \beta} ) \right].  
\end{align}
We note that although the structure factor has a slow $\cos (k\sigma)/(k\sigma)^2$ decay, the
integrand $ G_{\alpha\beta}(k_{\parallel}, L) + \frac{1}{2L}
S_{\alpha\beta}\left(k_{\parallel} \right) $ decays rapidly with
$k_{\parallel}$, making the numerical integration in Equation
(\ref{decay_integrate}) particularly easy. We also note that the integral over
$k_{\parallel}$ must be performed last since the divergent part needs to be
subtracted off by exploiting the asymptotic expansion of the difference between
a Riemann sum and the integral provided by the Euler-Maclaurin formula. Finally,
the force per unit area is obtained by numerically differentiating Equation (\ref{casimir_energy_msa}) with respect to $L$.

\begin{figure}
\centering
\subfigure[]{
\includegraphics[height=0.4\textwidth]{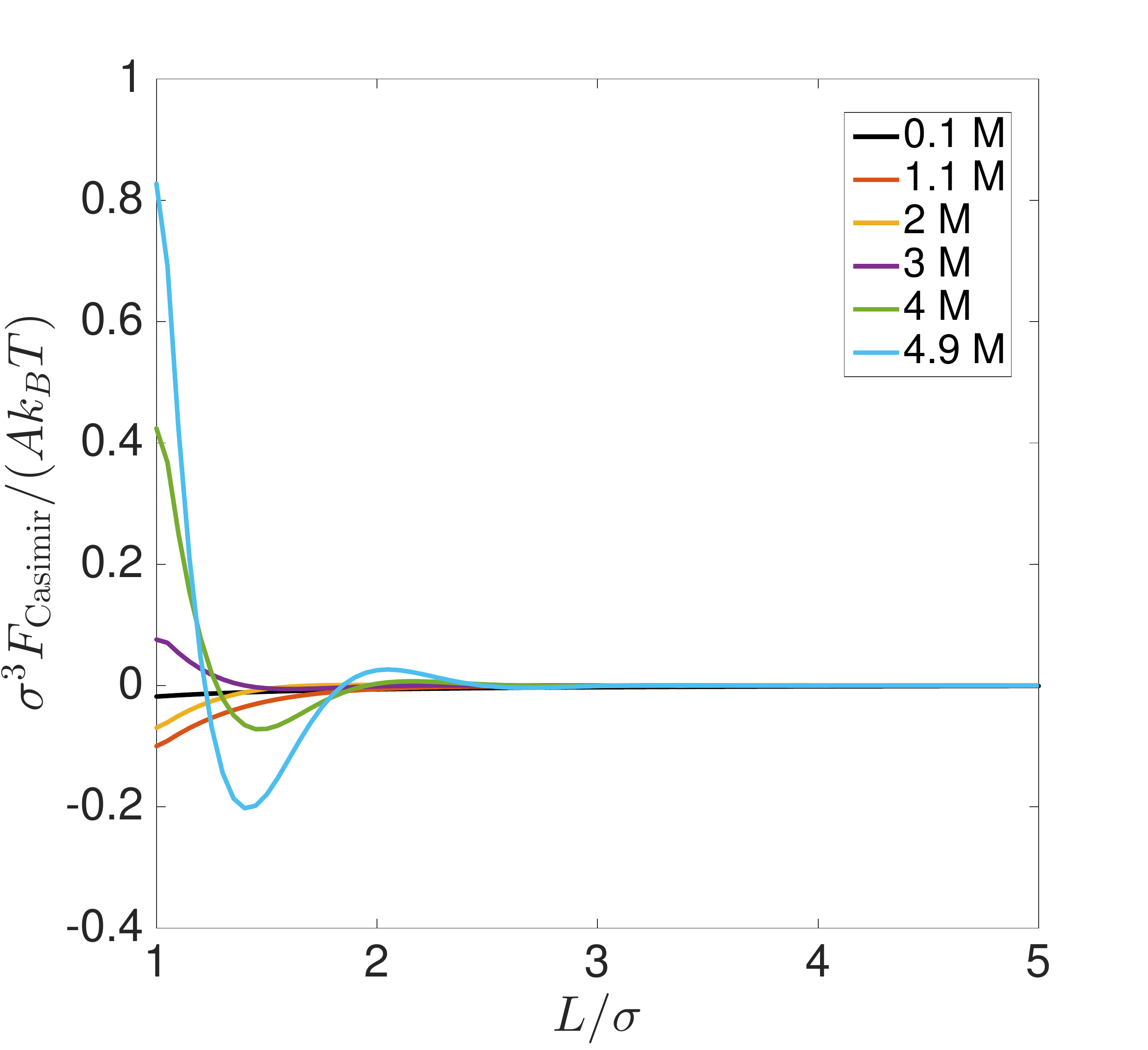}}
\subfigure[]{
\includegraphics[height=0.4\textwidth]{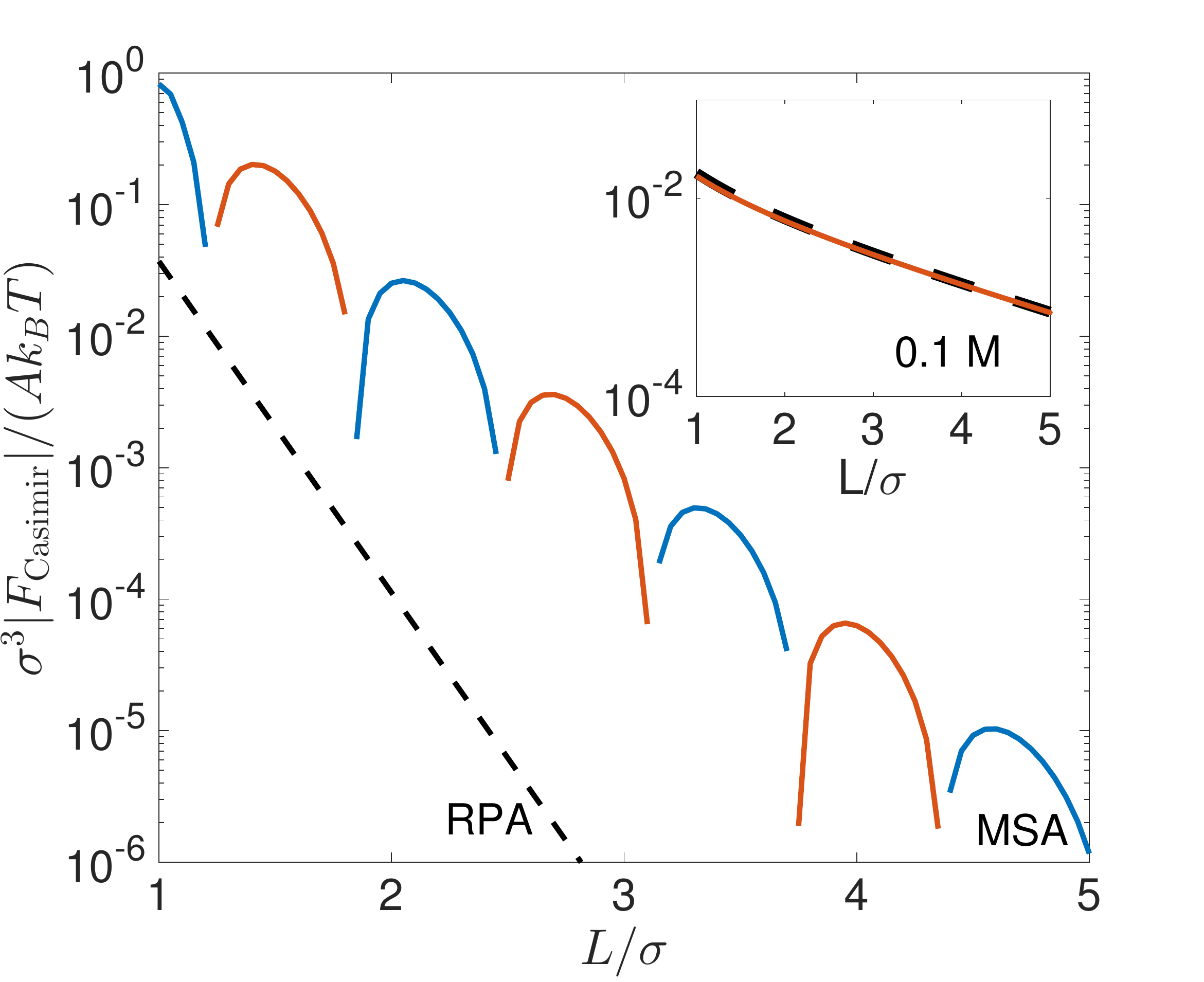}}
\caption{The electrolyte Casimir force for concentrated electrolytes oscillates
between attraction and repulsion as a function of surface separation due to hard
core repulsion. (a) The predicted electrolyte Casimir force for aqueous sodium
chloride solutions. The ion diameter and dielectric constant estimates are taken
from ref \cite{smith2016electrostatic}. (b) The main panel shows the electrolyte
Casimir force at 4.9M, a concentration close to saturation, plotted on a log
scale. The blue (red) portions denote repulsion (attraction), while the dashed
line indicates the RPA result at the same concentration.
The inset shows the Casimir force at 0.1M. }
\label{no_underscreening}
\end{figure}

As an illustration, we consider aqueous sodium chloride solutions, and use the ion
diameter and dielectric constant estimates from ref
\cite{smith2016electrostatic}. Figure \ref{no_underscreening} shows that the
predicted Casimir force as a function of surface separation is attractive for
low concentration, confirming the RPA result, but oscillates between
attraction and repulsion as a function of surface separation for concentrated
electrolytes. Figure \ref{no_underscreening}b shows that the decay length close
to saturation concentration is still comparable to the ion diameter, and at 4.9
M the screening length is $\approx 0.32 \sigma$, well below experimentally
measured values \cite{smith2016electrostatic}.   

\section{Conclusion}
\label{sec:conclusion}

We have used a second order expansion of the free energy of a binary ionic
liquid, confined between two charged insulating surfaces, in powers of the
fluctuating ion density modes, for a given spacing $L$ between the surfaces. The
resulting Casimir force acting between the surfaces is the derivative of this
free energy with respect to $L$ (cf. Equation~(\ref{casimir_force1})). 
The required input is provided by the partial structure factors
$S_{\alpha\beta}(k)$. We have examined two cases:
\begin{itemize}
\item[(a)] 
When the ions are assumed to be point charges, which amounts to the RPA,
valid for very low ion concentrations only, the calculations can be carried out
analytically, leading to the result in Equation~(\ref{casimir_force}); the Casimir force is
attractive, and decreases with a decay length equal to one half the Debye
length. This prediction agrees with earlier calculations based on a different,
fully microscopic Debye-H\"uckel approach~\cite{jancovici2004screening,
jancovici2005casimir,buenzli2005casimir}.
\item[(b)] 
At higher concentrations, finite size (excluded volume) effects become
predominant; we have included them within the MSA, which includes a short-range
contribution to the partial direct correlation functions, 
as shown in Equation~(\ref{MSA_dca}). 
The resulting expressions for the free energy and Casimir force must now
be evaluated numerically. The results for concentrated aqueous NaCl solutions,
within an implicit solvent model of oppositely charged hard spheres, are
summarized in Fig.~\ref{no_underscreening}. Instead of the exponential decay of the Casimir force
predicted by the RPA (point charges), the force now exhibits a striking,
exponentially damped oscillatory decay as a function of $L$ at the highest,
physically relevant concentrations. The periodicity of the oscillations is
comparable to the mean ion diameter, reflecting the structural ordering of the
ions. To the best of our knowledge, no such oscillatory Casimir force in electrolyte solutions has been
reported before, although oscillatory Casimir forces have been theoretically
predicted for active matter systems with a non-monotonic energy fluctuation
spectrum \cite{lee2017fluctuation}.
\end{itemize}
It must be stressed, however, that the Casimir force reported here is not
directly related to the ``underscreening" phenomenon discovered recently in
experiments~\cite{gebbie2013ionic,gebbie2015long,smith2016electrostatic,
lee2017scaling,lee2017underscreening}. Note that the ``first principles" theory of this 
phenomenon~\cite{rotenberg2017underscreening} is based on the same microscopic 
model and on the same theoretical tools employed in this paper.
The present calculations of the Casimir force can be readily extended to
asymmetric electrolytes (ions of different valences and diameters), as well as
to models of ionic solutions with explicit solvent~\cite{rotenberg2017underscreening},
within the same theoretical framework presented in Sections~\ref{sec:freeenergy}
and~\ref{sec:msa}. Work along these lines is in progress.
As a final remark, we note that the electrolyte fluctuation induced force discussed here
has to be considered even in the absence of a mean-field interaction arising from 
surface charges, and that other forces induced by surface-charge fluctuations
may also have to be taken into account under confinement by conducting walls
in or out of equilibrium~\cite{drosdoff_charge-induced_2016,dean_nonequilibrium_2016}.

\begin{acknowledgements}

B.R. acknowledges financial support from the French Agence Nationale de
la Recherche (ANR) under grant ANR-15-CE09-0013-01. A.A.L acknowledges support from the Winton Program for the Physics of Sustainability.  

This work is dedicated to Daan Frenkel on the occasion of his 70$^{th}$
birthday, as a token of appreciation of the authors' wonderful interactions
with him -- over a range of time-scales. In particular,
J.-P.H. wishes to express his sincere gratitude  
for an inspirational friendship and constant support over more than forty years.
\end{acknowledgements}




\end{document}